\documentstyle[prl,aps,epsf,twocolumn]{revtex}   
\begin{document}   
 
\title{ Phases of Wave Functions and Level Repulsion} 
  
\author{W.D. Heiss}

\address{Centre for Nonlinear Studies and
Department of Physics\\
 University of the Witwatersrand,
PO Wits 2050, Johannesburg, South Africa}

\maketitle


\begin{abstract}   
Avoided level crossings are associated with exceptional points
which are the singularities of the spectrum and eigenfunctions,
when they are considered as functions of a coupling parameter.
It is shown that the wave function of {\it one} state changes sign but not
the other, if the exceptional point is encircled in the complex plane.
An experimental setup is suggested where this peculiar phase change could
be observed.

\end{abstract}    

\vskip 1cm
   
The phase of the wave function changes in a characteristic 
way, if a selfadjoint Hamiltonian has an energy degeneracy at specific values
in some parameter space and if a loop around the point of degeneracy is
described in parameter space \cite{berry}. In the simple case of a
real symmetric Hamiltonian two parameters are needed to get a {\it diabolic
point}. In this case Berry's phase is $\pi $ when looping
around the degeneracy in two dimensional parameter space.

If a selfadjoint Hamiltonian depends only on one parameter, its variation
will in general give rise to level repulsion \cite{wig}. Associated with
a level repulsion is a pair of exceptional points \cite{kato} which are the 
points where the two levels actually coalesce when continued analytically
into the complex plane of the parameter \cite{hesa}. In the present letter
we discuss the behaviour of the wave functions and their phases when an
exceptional point is encircled. We suggest an experimental situation where 
such behaviour could be measured. It is distinctly different from a diabolic 
point where Berry's phase occurs. In particular, an exceptional point
must not be confused with a coincidental degeneracy of resonance states, 
which was considered in \cite{misiba} as a generalisation of Berry's phase.

The difference between a diabolic point and an exceptional point is due to
the selfadjointness of the Hamiltonian in the former and the lack of it in
the latter case. Also, when continuing into the complex parameter plane
we are faced with analytic functions of a complex variable which implies a
more rigid mathematical structure. In the references quoted above a
thorough discussion is given of the spectrum and eigenfunctions, when one
parameter on which the Hamiltonian depends is continued into the complex
plane. The parameter chosen can be an interaction strength as considered
in this paper, but other choices are possible \cite{bend}.
Generically, an $N$-dimensional matrix problem yields $N(N-1)$ exceptional
points. For an infinite dimensional problem, an infinite number can occur
\cite{he69}, but, depending on the particular situation, they may
accumulate at one point in the finite plane \cite{hemu}; there are special
cases, where no exceptional point occurs in the finite plane as can be seen 
from the analytic solution for the
single particle motion in the Hulthen potential \cite{newt}. 

\vspace{-2.6cm}

\begin{figure}
\epsfxsize=3.0in
\centerline{
\epsffile{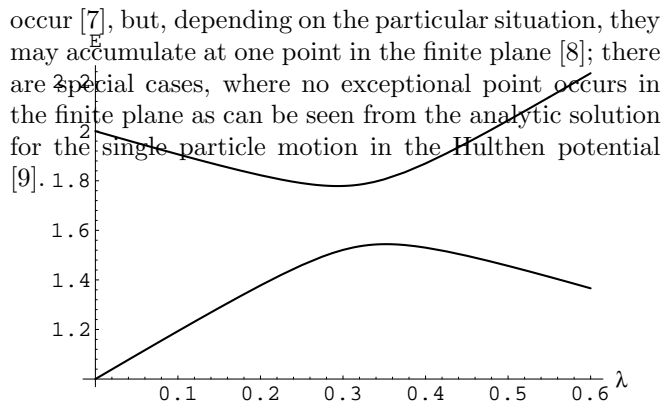}}
\vglue 0.15cm
\caption{
Level repulsion using in Eq.(\ref{eigv}) $\epsilon _1=1$, 
$\epsilon _2=2$, $\omega _1=2$, $\omega _2=-1$ 
and $\phi=\pi/25$.
}
\label{fig1}
\end{figure}

All essential aspects of exceptional points can be illustrated in a two
level model on an elementary level. In fact, for finite or infinite
dimensional problems an isolated exceptional point can be described locally
by a two dimensional problem \cite{hest}. In other words, even though a
high or infinite dimensional problem is globally more complex than the
two dimensional problem, we do not loose generality for our specific purpose 
when the restriction to a two dimensional problem is made. For easy
illustration we therefore confine ourselves to the discussion of
\begin{equation}  \label{ham}
H=\pmatrix {\epsilon _1 & 0 \cr 0 & \epsilon _2}+
\lambda U \pmatrix {\omega _1 & 0 \cr 0 & \omega _2} U^{\dagger}
\end{equation}
with
\begin{equation} \label{uang}
U=\pmatrix {\cos \phi & -\sin \phi \cr \sin \phi & \cos \phi }.
\end{equation}
This is, up to a similarity transformation, the most general form of a real
two dimensional Hamilton matrix of the type $H_0+\lambda H_1$. The particular
dependence on the parameter $\lambda $ has been chosen as it is of a nature
widely used in physical applications. We emphasise that our aim is not in
particular directed at a physical model that is describable by a two
dimensional problem although there may exist interesting problems in our
special context. The restriction has been chosen for illustration, while
the physical application that we have in mind is an infinite dimensional
situation.

\vspace*{-0.1cm}
\begin{figure}
\epsfxsize=3.0in
\centerline{
\epsffile{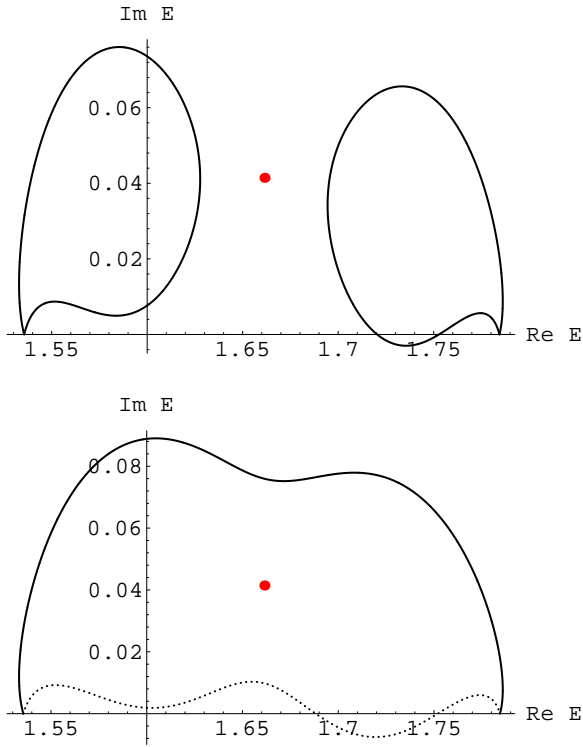}}
\vglue 0.15cm
\caption{
Contours in the complex energy plane. The top illustrates contours for
each level, if a closed loop is described in the $\lambda $-plane without
encircling the exceptional point. The bottom illustrates the energy
contour produced by two (equal) closed loops
in the $\lambda $-plane, which encircle the exceptional point. The solid
line corresponds to the first closed loop in the $\lambda $-plane and the 
dotted line to the subsequent loop. The solid dot is the position where the
contours would meet, if the loops would cross the exceptional point
in the $\lambda $-plane.}
\label{fig2}
\end{figure}

The eigenvalues of $H$ are given by
\begin{equation}  \label{eigv}
E_{1,2}(\lambda )={\epsilon _1+\epsilon _2+\lambda (\omega _1+\omega _2)
\over 2} \pm R
\end {equation}
where
\begin{eqnarray} \label{res}
R&=&\biggl\{({\epsilon _1-\epsilon _2\over 2})^2 \nonumber  \\
&+&({\lambda (\omega _1-\omega _2)\over 2})^2+{1\over 2}
\lambda (\epsilon _1-\epsilon _2)(\omega _1-\omega _2)\cos 2 \phi
\biggr\}^{1/2}.
\end{eqnarray}
Clearly, when $\phi =0$ the spectrum is given by the two lines
$$E_k^0(\lambda )=\epsilon _k+\lambda \omega _k, \quad k=1,2 $$
which intersect at the point of degeneracy
$\lambda =-(\epsilon _1-\epsilon _2)/(\omega _1-\omega _2)$. When the
coupling between the two levels is turned on by switching on $\phi $, the
degeneracy is lifted and avoided level crossing occurs as is illustrated
in Fig.1. Now the two levels
coalesce in the complex $\lambda $-plane where $R$ vanishes which happens
at the complex conjugate points
\begin{equation} \label{exc}
\lambda _c=-{\epsilon _1-\epsilon _2\over \omega _1-\omega _2}
\exp (\pm 2i\phi ).
\end{equation}

At these points, the two levels $E_k(\lambda )$ are connected by a square
root branch point, in fact the two levels are the values of one analytic
function on two different Riemann sheets. In Fig.2a we display
contours in the complex energy plane for each level, which are obtained
if a loop is described in the complex $\lambda $-plane, which does not
encircle the exceptional point. Correspondingly, Fig.2b shows the contours,
if the exceptional point is encircled. In this case only a double loop in
the $\lambda $-plane yields a closed loop in the energy plane.
Obviously, this connection is not of the type encountered at a genuine
diabolic point.

The difference has a bearing also on the scattering matrix \cite{mont}
and on the wave functions $\psi _1(\lambda )$
and $\psi _2(\lambda )$. In \cite{mont}, although the notion {\it exceptional
points} is not used, the pertinent distinction between a genuine degeneracy
of two resonances and the analytic coalescence (exceptional point) of two
(complex) eigenvalues is nicely discussed. A usual degeneracy of two
resonances still gives rise to a simple pole in the scattering matrix or
Green's function, while an exceptional point produces a double pole. With
regard to the eigenfunctions, we recall that for complex $\lambda $
the Hamiltonian
is no longer selfadjoint. This means that the eigenfunctions are no longer
orthogonal. We rather have a biorthogonal system which can be normalized as
\begin{equation} \label{bio}
\langle \tilde \psi _1(\lambda )|\psi _2(\lambda )\rangle =\delta _{1,2}
\end{equation}
where $|\psi _k\rangle $ and $\langle \tilde \psi _k|$ are the right hand and
left hand eigenvectors of $H$, respectively. Note that Eq.(\ref{bio}) causes
problems at the exceptional point, since it is exactly at this point where
two linearly independent eigenfuctions no longer exist. This is in contrast
to a genuine degeneracy of a selfadjoint operator where a $k$-fold degeneracy
always gives rise to a $k$-dimensional eigenspace. At the exceptional point,
not only the eigenvalues but also the eigenfunctions coalesce. As a
consequence, the orthogonality conflicts with the normalization, in other
words, if Eq.(\ref{bio}) is enforced globally, that is also at $\lambda =
\lambda _c$, the components of the wave function have to blow up. This can be
made explicit by parametrizing the wave functions by the complex angle
$\theta $, {\it viz.}
\begin{equation} \label{psi}
\psi _1(\lambda )=\pmatrix{\cos \theta \cr \sin \theta }, \quad
\psi _2(\lambda )=\pmatrix{-\sin \theta \cr \cos \theta }
\end{equation}
with
\begin{equation} \label{tan2}
\tan ^2\theta(\lambda )={
E_1(\lambda )-E_2(\lambda )-(\epsilon _1-\epsilon _2)-
\lambda (\omega _1-\omega _2)\cos 2\phi
\over
E_1(\lambda )-E_2(\lambda )+(\epsilon _1-\epsilon _2)+
\lambda (\omega _1-\omega _2)\cos 2\phi }.
\end{equation}
At $\lambda =\lambda _c$ it is $E_1=E_2$ and hence $\tan ^2\theta =-1$
which implies $|\cos \theta|=|\sin \theta|=\infty $, that is the components
of the wave functions blow up. (Note that $\tan ^2\theta \equiv 0$ in the
trivial case $\phi =0$). The increase of the components of the wave functions
while approaching exceptional points has been used in similar context as a
theoretical signature of a phase transition \cite{hemu}, but we do not
believe that it has observational consequences.

We now study the behaviour of the wave functions in more detail for two
contours in the $\lambda $-plane which start, say, at $\lambda =0$ and end
at large real values of $\lambda $, but enclose an exceptional point between
the two contours. For the complex angle $\theta $ we
choose an expression which is more convenient for this purpose, {\it viz.}
\begin{equation} \label{tan}
\tan \theta(\lambda )={\lambda (\omega _1-\omega _2)\sin 2 \phi
\over E_1(\lambda )-E_2(\lambda )+\epsilon _1-\epsilon _2
+\lambda (\omega _1-\omega _2)\cos 2 \phi }.
\end{equation}
The first path can be taken along the real $\lambda $-axis. From Eq.(\ref{tan})
we read off the expected result that $\theta (0)=0$ and $\theta (\lambda )\to
\phi $ for $\lambda \gg |(\epsilon _1-\epsilon _2)/(\omega _1-\omega _2)|$.
In obtaining this result use is made of
$E_1-E_2=2R\to \lambda (\omega _1-\omega _2)$
for $\lambda \gg |(\epsilon _1-\epsilon _2)/(\omega _1-\omega _2)|$. 
For the second path we move
into the upper $\lambda $-plane in order to pass above the exceptional point
before returning down to the real axis again. Using again Eq.(\ref{tan}) we
now have to observe that we crossed into the other sheet which means 
$E_1-E_2=-2R\to -\lambda (\omega _1-\omega _2)$. As a consequence we find
this time $\tan \theta=-\cot \phi=\tan (\phi +\pi/2 )$. Surely we would expect
the wave functions to interchange just like the energies, if an exceptional
point is encircled. But our finding indicates that {\it one wave function
has changed its sign}. In fact, we obtain along the second path
\begin{equation} \label{psiph}
\psi _1\to \psi _2, \quad \psi _2\to -\psi _1 .
\end{equation}
Note that this result is equally obtained, if a closed contour which encloses
the exceptional point is described in the $\lambda $-plane. Of interest
is the result of a double loop in the $\lambda $-plane. It yields
\begin{equation} \label{berph}
\psi _1\to -\psi _1, \quad \psi _2\to -\psi _2 
\end{equation}
which is, in accordance with the corresponding single loop in the energy
plane, just Berry's phase retrieved. We mention that this implies
that the wave functions, if parametrized as in Eq.(\ref{psi}), have an
algebraic singularity that is determined by a fourth root at the exceptional
point; only a four-fold loop in the $\lambda $-plane restores completely
the original situation.

Next we address the question concerning the physical significance of these
findings. It boils down to the problem of varying a {\it complex} interaction
parameter in the laboratory. Problems of a similar nature have been discussed
in connection with Berry's phase for dissipative systems \cite{NeKeEl}, but
not as yet for exceptional points. We are guided by the phenomenological
description of open quantum systems \cite{hemu} and submit as one suggestion
a strongly absorptive system, where the parameter $\lambda $ in Eq.(\ref{ham})
is traditionally replaced by $-iG$ with real absorption parameter $G$. 
With the replacement the eigenvalues acquire imaginary parts which are
related to the inverse life times of the states of the open system.
The exceptional points appear now in the complex $G$-plane at
\begin{equation} \label{excab}
G_c=-{\epsilon _1-\epsilon _2\over \omega _1-\omega _2}
\exp (\pm 2i\phi +i\pi/2).
\end{equation}
If the coupling of the two channels is equal, that is if $\phi=\pi /4$, the
two exceptional points lie on the real $G$-axis at
$$G_c=\pm {\epsilon _1-\epsilon _2\over \omega _1-\omega _2}.$$
If the coupling is nearly equal, they lie just above or below the real axis
depending on the coupling being slightly weaker or stronger ($\phi <\pi/4$
or $\phi >\pi/4$).
Controlling the absorption parameter $G$ {\it and} the relative coupling
enables one to pass the exceptional point on a path below and above, which is
the situation described above. A setup similar to the one proposed by Berry
\cite{berry} and implemented in \cite{wedub}
should make feasible a distinction of the two cases. In a
two dimensional electromagnetic resonator the relative coupling of two 
suitable levels, that is levels that do coalesce, can be controlled by
a judicious choice of the geometry of the resonator \cite{wedub}. The global 
absorption can be controlled
by radiation losses regulated by suitable antennas. In this way the change
of the phase of {\it one but not the other} wave function should be 
observable. Note that, for large values of $G$, one state will be much 
broader than the other \cite{hemu}.

We stress again that the occurrence of exceptional points is a generic
mathematical feature associated with any system that has avoided level
crossing. While this presentation used a two dimensional illustration,
the results can be immediately generalised. In fact, if more than one
exceptional point is encircled, the resulting phase change is simply a
combination of several two--dimensional cases. The
findings presented in this paper have therefore universal significance.
Whether the particular phase changes of the respective wave functions are
detectable, is subject to experimentation. Due to the short life time
involved for one of the states, scattering experiments, where a steady
incoming flux can be maintained, are expected to be particularly suitable.
Fortuitously, in the spirit of the experiment performed by \cite{wedub} the
adiabatic change of the parameters is not essential, since it is the movement 
of the nodes of the wave functions that are being observed under parameter 
variation \cite{darm}. Possible jumps of the phases do therefore not affect 
the result of interest.

\end{document}